\documentclass[fleqn,twoside]{article}
\usepackage{espcrc2}
\usepackage{graphicx}
\usepackage{epsf}

\title{Stability of two-fermion bound states in the explicitly
covariant Light-Front
Dynamics}

\author{M. Mangin-Brinet,\address{Institut des Sciences
Nucl\'eaires, \\ 
        53, avenue des Martyrs, 38 026 Grenoble, France}%
J. Carbonell$^{\mbox{{\small a}}}$        
and
V. A. Karmanov\address{Lebedev Physical Institute,\\ 
Leninsky Pr. 53, 119991 Moscow, Russia}}

\begin{document}

\maketitle

\section{Introduction}
The system of two bound fermions covers a huge number of
interesting
problems from atomic, nuclear and subnuclear physics. It is one
of the
most difficult problems in field theory due to the fact that
bound states
necessarily involve an infinite number of diagrams. We studied
this
problem in the framework of the explicitly covariant light-front
dynamics \cite{karm76}
(CLFD). In this approach, the state vector is defined on an
hyperplane given 
by the invariant equation $\omega\cdot x=0$ with
$\omega^2=0$. The standard light-front, 
reviewed in \cite{BPP_PR_98}, is recovered for
$\omega=(1,0,0,-1)$. 
The CLFD equations have been solved exactly for a two fermion 
system with different boson exchange ladder kernels
\cite{These_MMB,fermions}.
We have considered separately the usual couplings between two
fermions 
(scalar, pseudo-scalar, pseudo-vector, and vector)
and we were interested in states
with given angular momentum and parity $J^\pi=0^{\pm},1^{\pm}$. 
Each coupling leads to a system of integral equations, which in
practice are solved 
on a finite momentum domain $[0; k_{max}]$. If the solutions
necessarily exist when 
the integration domain is finite -- for the kernels are compact,
it is not a
priori obvious that the equations admit stable solutions when
$k_{max}$ goes to infinity. Particular attention must therefore 
be paid to the stability of 
the equations relative to the cutoff $k_{max}$. We develop
hereafter an analytical 
method to study the cutoff dependence of the
equations and to determine whether they need to be regularized
or not. 
 
The method  will here be detailed for a $J=0^+$ state 
in the Yukawa model but it can be applied to any coupling. 
Results will be presented for scalar and pseudo-scalar 
exchange. This latter furthermore exhibits some strange
particularities which will be discussed.

\section{Scalar exchange}

Let us consider a system of two fermions in a $J^{\pi}=0^+$
state, bound by a
scalar exchange, whose Lagrangian density
is given by ${\cal{L}}=g_s\bar{\Psi}\Phi\Psi$. 
Its wave function, constructed using all possible 
spin structures, is determined in the $0^+$ case by two
components \cite{ckj0},
$f_1$ and $f_2$, which depend on the two scalar variables $k$
and
$\theta$:
\begin{eqnarray*}
\psi={1 \over \sqrt{2}}w^{\dag}_{\sigma_2}
\left( f_1+i{\vec{\sigma} \cdot [\vec{k}\times \hat{n} ] \over 
k\sin\theta} f_2\right) \sigma_y
w^{\dag}_{\sigma_1}
\end{eqnarray*}
$\vec{k}$ is the momentum of one particle in the system of
reference where $\vec{k}_1+\vec{k}_2=0$, $\hat{n}$ is the
spatial part
of the normal $\omega$ to the light-front plane, $\theta$ is the
angle 
between $\vec{k}$ and $\hat{n}$, and $w$ is the two component
spinor. The
appearence of a second component compared to the non
relativistic
case is due to vector $\hat{n}$, which induces additional 
spin structures. 

$f_1$ and $f_2$ satisfy the system of coupled equations :
\begin{eqnarray}\label{eqfiJ0}  
&&\left[M^2-4(k^2 +m^2)\right]f_1(k,\theta)\nonumber \\ 
&=&\frac{m^2}{2\pi^3} \int
\left[K_{11}
f_1(k',\theta')+K_{12}
f_2(k',\theta')\right]\frac{d^3k'}{\varepsilon_{k'}}
\nonumber\\
&&\left[M^2-4(k^2 + m^2)\right] f_2(k,\theta) 
\nonumber\\
&=&\hspace{-0.3cm}\frac{m^2}{2\pi^3} \int \left[K_{21}
f_1(k',\theta')+K_{22}
f_2(k',\theta')\right]\frac{d^3k'}{\varepsilon_{k'}}
\end{eqnarray}
$M^2$ si the total mass squared of the system, $m$ is the
constituent 
mass and $\varepsilon_k=\sqrt{\vec{k}^2+m^2}$. 
The kernels $K_{ij}$ result from a first integration of more
elementary 
quantities:
\begin{eqnarray*} 
K_{ij}(k,\theta;k',\theta')&=&\int_0^{2\pi}{\kappa_{ij} \over
(Q^2+\mu^2)m^2\varepsilon_k \varepsilon_{k'}}{d\varphi'\over
2\pi}, 
\end{eqnarray*} 
where $\kappa_{ij}$ depend on the type of coupling.
The analytical expressions of $\kappa_{ij}$ for the scalar
coupling, read
\begin{eqnarray}\label{eqap1}
\kappa^{S}_{11}\hspace{-0.2cm}&=&\hspace{-0.2cm}-\alpha\pi
\left[2 k^2 k'^2+3k^2 m^2+3k'^2 m^2+4 m^4 \right. \nonumber \\
\hspace{-0.2cm}&&\hspace{-0.2cm}-2 k k'\varepsilon_k \varepsilon_{k'} \cos\theta \cos\theta'
\nonumber\\
\hspace{-0.2cm}&&\hspace{-0.2cm}\left.- k k' (k^2 + k'^2 + 2 m^2)
\sin\theta \sin\theta' \cos\varphi'\right]\nonumber\\
\kappa^{S}_{12}\hspace{-0.2cm}&=&\hspace{-0.2cm}-\alpha\pi m
(k^2 - k'^2) \left(k'\sin\theta' + k\sin\theta\cos\varphi'
\right)
\nonumber\\
\kappa^{S}_{21}\hspace{-0.2cm}&=&\hspace{-0.2cm}-\alpha\pi m
(k'^2 - k^2) \left(k\sin\theta + k'\sin\theta'\cos\varphi'
\right)
\nonumber\\
\kappa^{S}_{22}\hspace{-0.2cm}&=&\hspace{-0.2cm}-\alpha\pi
\left[\left(2 k^2 k'^2+3k^2 m^2+3k'^2 m^2+4 m^4 \right.
\right.\nonumber \\
\hspace{-0.2cm}&&\hspace{-0.2cm}\left. - 2 k k' \varepsilon_k\varepsilon_{k'}
\cos\theta \cos\theta'\right)\cos\varphi' 
\nonumber\\
\hspace{-0.2cm}&&\hspace{-0.2cm}\left.-k k'(k^2 + k'^2 + 2 m^2) \sin\theta\sin\theta'\right]
\end{eqnarray}
In practice, the integration region over the momenta is reduced
to a finite
domain $[0,k_{max}]$. The kinematical term $[M^2-4(k^2+m^2)]$
on l.h.s. of equation (\ref{eqfiJ0}) does not
generate any singularity and the kernels $K_{ij}$ 
are smooth functions
of the $\theta$ variable. Thus, the stability of the solution 
depends only on
the asymptotical behavior of the kernels in the $(k,k')$ plane.

Variables $(k,k')$ can tend to infinity following different
directions:  for a 
fixed value of $k$, $K_{11}$ decreases as $1/k'$, and vice
versa. As the
integration volume contains the factor $\varepsilon_{k'}$, this
means 
that the total kernel decreases as $1/k'^2$, that is like a
Yukawa 
potential. In contrast, $K_{22}$ does not decrease in any
direction of
the $(k,k')$ plane, but tends to a positive constant 
with respect to $k$ and $k'$. $K_{22}$ 
is thus asymptotically repulsive and does not generate any
unstability. In the domain
where both $k,k'$ tend to infinity with a fixed ratio ${k'\over
k}=\gamma$, it is useful to introduce the functions $A_{ij}$
defined by
\begin{eqnarray*}
K_{ij}=-\frac{\pi\alpha}{m^2}\left\{
\begin{array}{ll}
\sqrt{\gamma}      \;A_{ij}(\theta,\theta',\gamma)   & \mbox{if
$\gamma\leq 1$}\\
{1\over\sqrt\gamma}\;A_{ij}(\theta,\theta',1/\gamma) & \mbox{if
$\gamma\geq 1$}
\end{array}\right.
\end{eqnarray*}
Since $K_{22}$ is repulsive and does not generate any collapse,
we
consider only the first channel. We have 
\begin{eqnarray*}
&&\hspace{-0.5cm}A_{11}(\theta,\theta',\gamma)=
\frac{1}{\sqrt{\gamma}}\int_0^{2\pi}\frac{d\varphi'}{2\pi}
{1\over D} \times\\ 
&&\hspace{-0.5cm}\left\{2\gamma(1-\cos\theta\cos\theta')-   
(1+\gamma^2)\sin\theta\sin\theta'\cos\varphi'\right\}  
\end{eqnarray*}
where  
\begin{eqnarray*}
D&=&(1+\gamma^2)(1+|\cos\theta-\cos\theta'|-\cos\theta\cos\theta')\\
&-&2\gamma\sin\theta\sin\theta'\cos\varphi'
\end{eqnarray*}
Let us now majorate the function $A_{11}$. For fixed $\gamma$, 
the maximum of $A_{11}$ is achieved at $\theta=\theta'$ 
and for any $\theta=\theta'$ it reads: 
$A_{11}(\theta=\theta',\gamma)=\alpha'\sqrt{\gamma}$. The 
maximum value of kernel $K_{11}$ is thus reached for $\gamma=1$.
The majorated kernel obtained this way coincides with the
non-relativistic potential $U(r)=-\alpha'/r^2$ in the momentum
space  
with $\alpha'=\alpha/(2m\pi)$.
As well known \cite{ll}, for this potential,
the binding energy does not depend on cutoff if
$\alpha'<\alpha_{cr}=1/(4m)$
what restricts the coupling constant to:
$\alpha<\pi/2$. If $\alpha'>1/(4m)$, the binding 
energy is cutoff dependent and tends to $-\infty$ when
$k_{max}\to \infty$. 
A finer majoration of $A_{11}$ was done by  
taking into account its dependence on $\gamma$ \cite{mck_prd1}. 
In this way we have found  $\alpha_{cr}=\pi$, instead of
$\pi/2$. 
As the kernel was majorated, the critical coupling constant is
expected to
be larger than $\pi$. 

\begin{figure}[htbp]
\vspace{-1.5cm}
\begin{center}
\epsfxsize=7.5cm\epsfysize=7.5cm\mbox{\epsffile{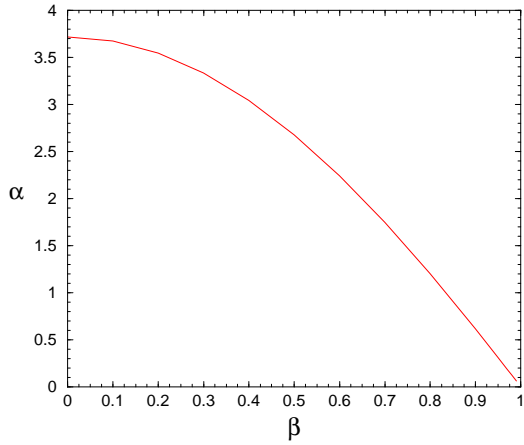}}
\end{center}
\vspace{-1.2cm}
\caption{Function $\alpha(\beta)$ for LFD Yukawa model with
$K_{11}$ channel only.}
\label{alpha_beta} 
\vspace{-0.3cm}
\end{figure}

It can be determined, together with the asymptotical
behavior of the wave functions, by considering  the limit
$k\to\infty$ of
equation (\ref{eqfiJ0}) for $f_1$
\begin{eqnarray*}
&&-4 f(k,z)=  \\
&& \frac{m^2}{\pi^2} 
\int_0^{\infty} \gamma d\gamma \int_{-1}^{+1}dz' 
K(k,z;\gamma k,z')f(\gamma k,z') 
\end{eqnarray*}
where we have neglected the binding energy, supposing that it is
finite, and omitted the indices
for $f_1$ and $K_{11}$. This can also be written
\begin{eqnarray}\label{as2_1}  
4f(k,z)&=&{\alpha\over\pi} \int_{-1}^1 dz' 
 \int_0^1  d\gamma\; A(\theta,\theta',\gamma) \nonumber\\
&\times&\hspace{-0.3cm}\left\{\gamma^{3/2} f(\gamma
k,z')+\gamma^{-5/2} f({k\over\gamma},z')\right\} 
\end{eqnarray}
Looking for a solution which behaves as 
\begin{equation}\label{pl}
f(k,z) \sim {h(z)\over k^{2+\beta}},  \qquad 0\le\beta< 1. 
\end{equation}
we are led for $h(z)$ to the eigenvalue equation
\begin{eqnarray*} 
h(z)=\alpha\int_{-1}^{+1} dz' H_{\beta}(z,z')\;h(z')  
\end{eqnarray*}
with
\begin{eqnarray*} 
H_{\beta}(z,z')=\int_0^1{d\gamma\over2\pi\sqrt\gamma} 
A(z,z',\gamma)\,\cosh{(\beta\log\gamma)} 
\end{eqnarray*} 
The relation 
between the coupling constant $\alpha$ and the coefficient
$\beta$, 
determining the power law  of the asymptotic wave function, can
be found in 
practice by solving the eigenvalue equation (\ref{Sta_5}) for a
fixed value 
of $\beta$
\begin{equation}\label{Sta_5}
\lambda_{\beta} \, h(z)=\int_{-1}^{+1} dz'
H_{\beta}(z,z')\;h(z')  
\end{equation}
and taking $\alpha(\beta) =1/\lambda_{\beta}$
The relation $\alpha(\beta)$ obtained that way is represented in
Figure
\ref{alpha_beta}. The value $\beta=0$ corresponds to the maximal
-- that is the critical -- value of $\alpha$: 
$\alpha_c=\alpha(\beta=0)=3.72$, in agreement with the previous 
analytical estimations. It is independent of the exchanged mass
$\mu$.

\begin{figure}[htbp]
\begin{center}
\vspace{-1.4cm}
\epsfxsize=7.5cm\epsfysize=7.2cm\mbox{\epsffile{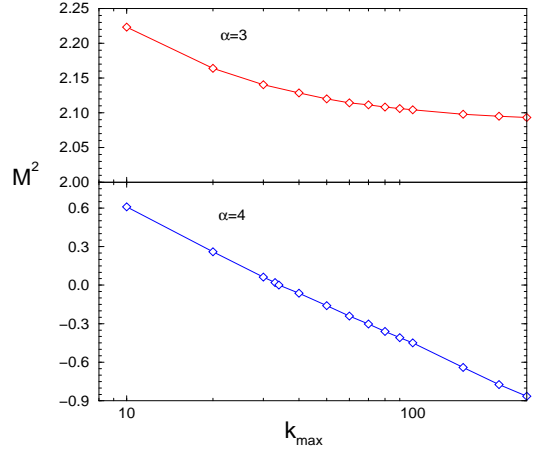}}
\end{center}
\vspace{-0.8cm}
\caption{Cutoff dependence of the binding energy in the $J=0^+$ 
state ($\mu=0.25$), in the one-channel problem ($f_1$), for two fixed values
of the coupling 
constant below and above the critical value.}\label{B_kmax}
\vspace{-0.3cm}
\end{figure}
Figure \ref{B_kmax} shows the two different regimes, whether the
coupling 
constant is
below $(\alpha=3)$ or above $(\alpha=4)$ the critical value
$\alpha_c$.
\begin{figure}[htbp]
\vspace{-1.9cm}
\begin{center}
\epsfxsize=7.5cm\epsfysize=7.5cm\mbox{\epsffile{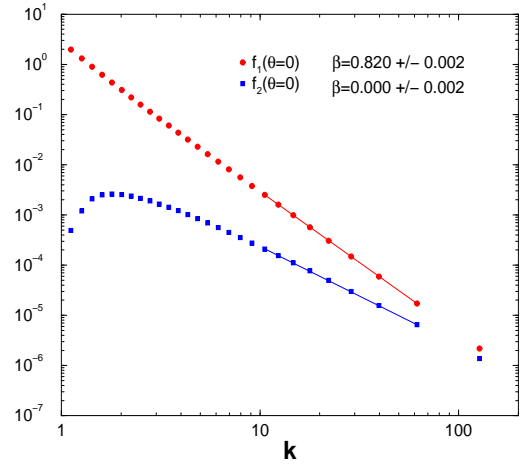}}
\end{center}
\vspace{-1.0cm}
\caption{Asymptotical behavior of the $J=0^+$ wave function 
components $f_i$ for $B$=0.05, $\alpha$=1.096, $\mu$=0.25. The
slope coefficient are $\beta_1=0.82$ and  $\beta_2\approx0$.}
\label{wf_as_50} 
\vspace{-0.6cm}
\end{figure}
As it can be seen in Figure \ref{wf_as_50}, the wave functions
accurately follow the power law
asymptotical behavior $1/k^{2+\beta}$ with a coefficient 
$\beta(\alpha)$ given in Figure \ref{alpha_beta}. 
It is worth noticing that -- at least in the framework of this
model --
one could measure the coupling constant
from the asymptotic behavior of the bound state wave function.

A similar study has been done for the $J=1^+$ state, which is
shown to be unstable
without regularization \cite{mck_prd1,mck_prd2}.

%%%%%%%%%%%%%%%%%%%%%%%%%%%%%%%%%%%%%%%%%%%%%%%%%%%%%%%%%%%%%%%%%%%%%%%%%%
\section{Pseudo-scalar coupling}

The stability of the pseudo-scalar (PS) coupling is analyzed
similarly to
the scalar one. The same method leads to the conclusion that 
the equations for the PS coupling are quite surprisingly 
stable without any regularization.

\begin{figure}[htbp]
\begin{center}
\vspace{-1.5cm}
\mbox{\epsfxsize=7.5cm\epsfysize=7.5cm\epsffile{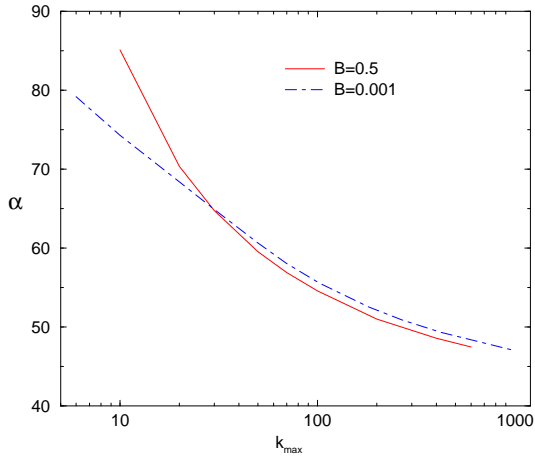}}
\vspace{-0.9cm}
\caption{Convergence of coupling constants as function of the
cutoff $k_{max}$ for $B=0.001$ and $B=0.5$. The exchange mass is $\mu=0.15$.}
\label{PS_total1_500}
\end{center}
\vspace{-0.7cm}
\end{figure}

However, the results show a "quasi-degeneracy" of the coupling
constant,
for a wide range of binding energies. One has for instance 
(see Figure \ref{PS_total1_500}) $\alpha=49.5$ 
for a system with $B=0.001$, and $\alpha=48.6$ for a system five
hundred 
times more deeply bound ($B=0.5$), that is only a 2\%
difference. 

This peculiar behavior can be shown to come from the second
channel:
\begin{eqnarray}\label{lfd_K22}
&&[M^2-4(k^2+m^2)] f_2(k,\theta)= \nonumber \\
&&{m^2\over 2\pi^3} \int 
K^{PS}_{22}(k,k',\theta,\theta',M^2)f_{2}(k',\theta'){d^3k'\over
\varepsilon_{k'}}
\end{eqnarray}
\par

\begin{figure}[htbp]
\vspace{-1.5cm}
\begin{center}
\begin{minipage}{40mm}
\vspace{0.5cm}
\hspace{-2.0cm}
\mbox{\epsfxsize=4.0cm\epsfysize=4.0cm\epsffile{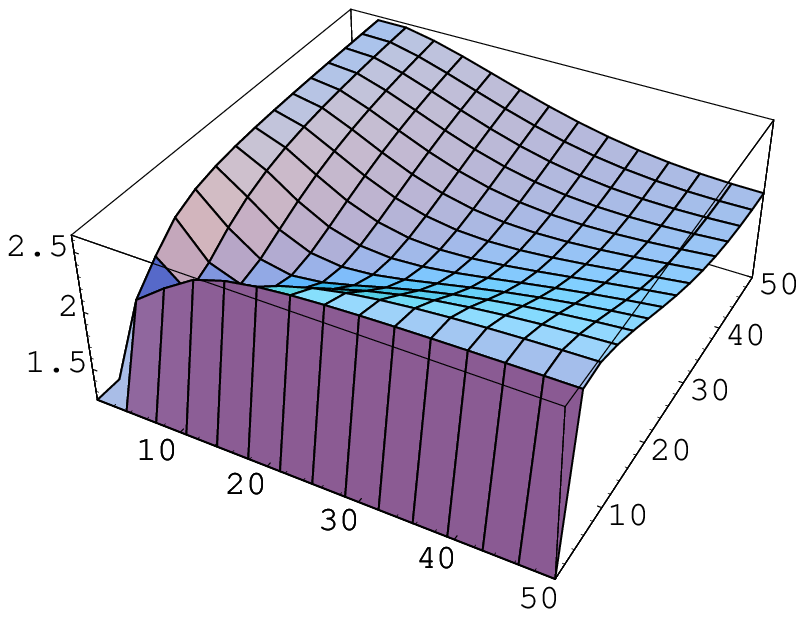}}
\end{minipage}\hspace{1.5cm}
\begin{minipage}{50mm}
\vspace{-4.0cm}
\hspace{2.3cm}
\mbox{\epsfxsize=4.0cm\epsfysize=3.5cm\epsffile{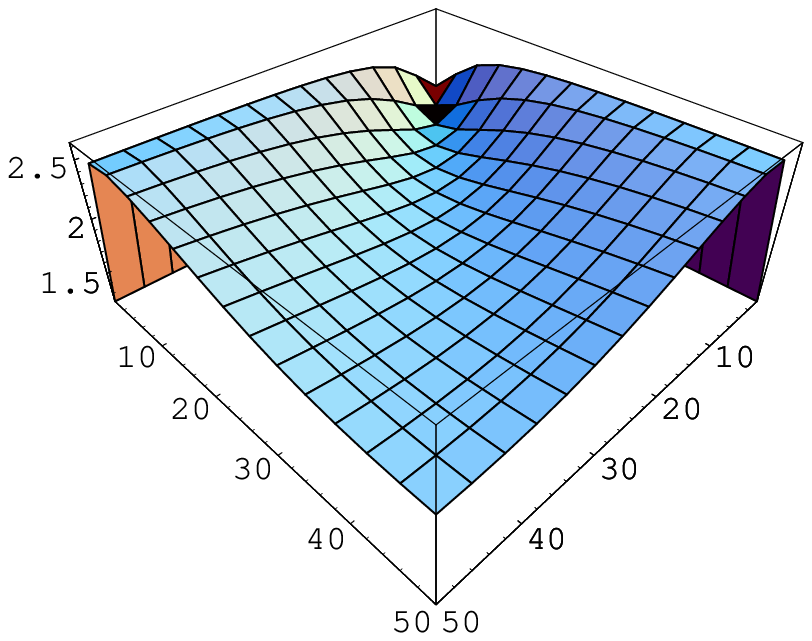}}
\end{minipage}
\vspace{-0.5cm}
\caption{$K_{22}$ kernel in $(k,k')$ plane.}\label{K22PSplot}
\end{center}
\vspace{-1.0cm}
\end{figure}

The kernel $K_{22}(k,k',\theta,\theta',M^2)$, whose expression 
is explicitly given in \cite{These_MMB}, is represented in 
Figure \ref{K22PSplot} for fixed values of $\theta,\theta'$. It 
vanishes for 
$k=0$ or $k'=0$ and tends towards a positive constant in all the
$(k,k')$ plane. 

\begin{figure}[htbp]
\vspace{-1.5cm}
\begin{center}
\begin{minipage}{40mm}
\vspace{0.5cm}
\hspace{-2.0cm}
\mbox{\epsfxsize=4.0cm\epsfysize=4.0cm\epsffile{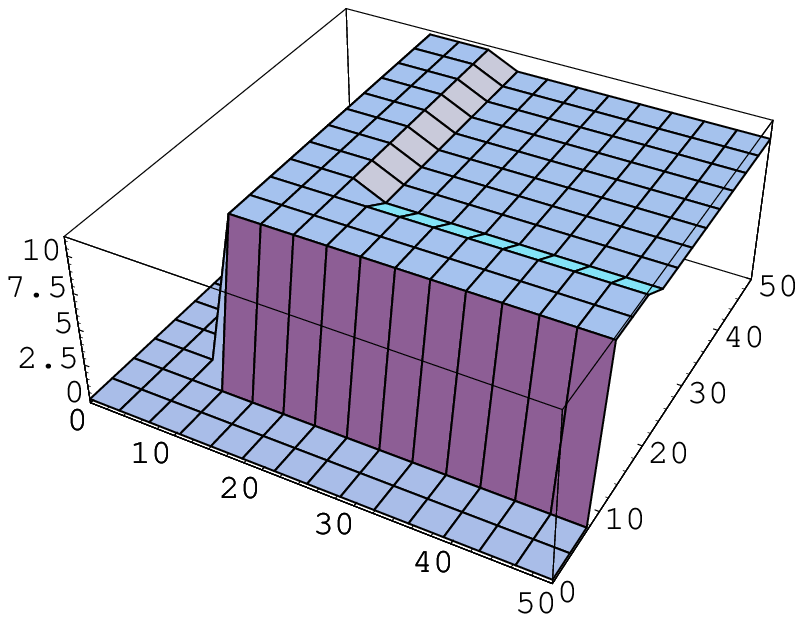}}
\end{minipage}\hspace{1.5cm}
\begin{minipage}{50mm}
\vspace{-4.0cm}
\hspace{2.3cm}
\mbox{\epsfxsize=4.0cm\epsfysize=3.5cm\epsffile{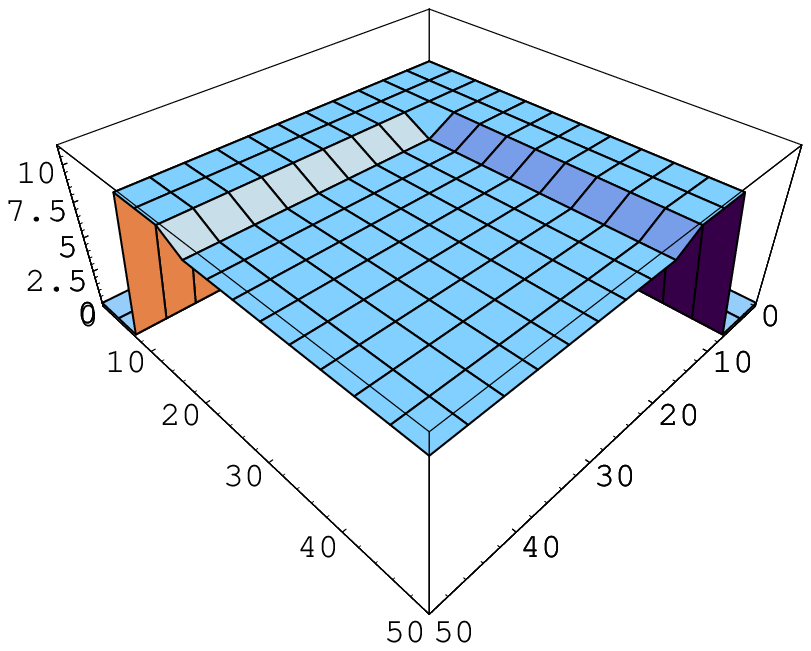}}
\end{minipage}
\vspace{-1.0cm}
\caption{Modelization of $K_{22}$ by a simpler kernel in the
$(k,k')$
plane.}\label{pot_modele_PS}
\end{center}
\vspace{-1.0cm}
\end{figure}

Let us modelize this kernel by a kind of "potential barrier" in
the momentum space ($k,k'$), displayed in Figure
\ref{pot_modele_PS}, whose advantage is to be analytically
solvable. 
\begin{eqnarray*}
&\hspace{-0.3cm}&K(k,k')= {\alpha U_1\over
m^2}\Theta(k'-k_1)\Theta(k_2-k') \\
&\hspace{-0.3cm}\times&\hspace{-0.3cm} 
\left[\Theta(k-k_1)\Theta(k_2-k)+\Theta(k-k_2)\Theta(k_{max}-k)\right]\\
&\hspace{-0.3cm}+&\hspace{-0.3cm}
\Theta(k'-k_2)\Theta(k_{max}-k')\left[{\alpha U_1\over m^2}
\Theta(k-k_1)\right.\\
&\hspace{-0.3cm}\times&\hspace{-0.3cm} \left.\Theta(k_2-k)+
{\alpha U_2\over m^2}
\Theta(k-k_2)\Theta(k_{max}-k)\right]
\end{eqnarray*}
with $\Theta(x)=1,\,x>0$ and $\Theta(x)=0, \; x\le0$.
This kernel has the same characteristics than $K^{PS}_{22}$ 
since it is zero when $k,k'\to 0$,and tends towards a constant
when
$(k,k')$ go to infinity with a fixed ratio $\gamma=k'/k$. 

$f_2$ satisfies the Schr\"odinger type equation
\begin{eqnarray}\label{modele_PS}
[k^2+\kappa^2] f_2(k)=\hspace{-0.1cm}-{m^2\over (2\pi)^3} \int 
K(k,k')f_2(k'){d^3k'\over k'}\hspace{-0.17cm}
\end{eqnarray}
with $\kappa^2=m^2-{M^2\over 4}$.
We assume that $k_1<k_2<k_{max}$ et $U_2<U_1$.
The term $\varepsilon_{k'}$ in the volume element of
(\ref{lfd_K22}) 
was replaced by its large momentum behavior, that is by $k'$. 
We define $\Gamma(k)=[k^2+\kappa^2] f_2(k)$. The equation for 
$\Gamma(k)$, which is analytically solvable, reads:
\begin{eqnarray*} 
\Gamma(k)=-{m^2\over 2\pi^2} \int_{-\infty}^{+\infty} 
K(k,k'){\Gamma(k')\over k'^2+\kappa^2} k'dk'
\end{eqnarray*}
The solution $\Gamma(k)$ is constant for $k_1<k<k_2$ and
$k_2<k<k_{max}$ :
\begin{eqnarray*}
\Gamma(k)&=&\Gamma_1 \Theta(k-k_1) \Theta(k_2-k) \\
&+&\Gamma_2 \Theta(k-k_2) \Theta(k_{max}-k)
\end{eqnarray*}
The $\Gamma_i$ satisfy the coupled equations
\begin{eqnarray*}\label{syst_G1G2}
\left\{
\begin{array}{ccc}
(1+\alpha u_1 a)\;\Gamma_1&=& -\alpha u_1 b\Gamma_2     \\
(1+\alpha u_2 b)\;\Gamma_2&=& -\alpha u_1 a\Gamma_1     
\end{array}
\right.
\end{eqnarray*}
where we have defined $u_i={m^2\over 2\pi^2}U_i$ and
\begin{equation}\label{ab}
a=\log\left(k_2^2+\kappa^2 \over k_1^2+\kappa^2\right), \quad
b=\log\left(k_{max}^2+\kappa^2 \over k_2^2+\kappa^2\right)  
\end{equation}
Replacing $\Gamma(k)$ by its definition in terms of $f_2(k)$,
we finally get the solution of equation (\ref{modele_PS}) on the
form:
\begin{eqnarray*}
f_2(k)&=& {N\over
k^2+\kappa^2}\left[\Theta(k-k_1)\Theta(k_2-k)\right. \\
&-&\left.{\alpha a u_1 \over (1+\alpha b u_2)} 
\Theta(k-k_2)\Theta(k_{max}-k)\right]
\end{eqnarray*}
where $N$ is a normalisation constant.
For a given $\kappa$ the coupling constant is
\begin{eqnarray*}
\alpha(\kappa)={(au_1+bu_2)+\sqrt{(au_1-bu_2)^2+4u_1^2ab}\over 
2abu_1(u_1-u_2)}
\end{eqnarray*}
and the results provided by this simple kernel 
are summarized in Table \ref{alphamodel2}.
\begin{table}[htbp]\caption{Coupling constant as a function of
the binding energy 
for the analytical model of the PS coupling second
channel.}\label{alphamodel2}
\[
\begin{array}{||c|c||}\hline
B              & \alpha \\\hline
  0.001        & 22.5327        \\\hline
  0.010    & 22.5334    \\\hline
  0.100    & 22.5396    \\\hline
  0.500    & 22.5639    \\\hline
  1,00     & 22.5862    \\\hline
  1.50     & 22.5995    \\\hline
  2.00     & 22.6040    \\\hline
\end{array}
\]
\vspace{-1.0cm}
\end{table}
$\alpha(\kappa)$ depends on $\kappa$ through logarithms in 
$a,b$, eq. (\ref{ab}). Besides, 
the value of $\kappa$ is much smaller than $k_1,k_2,k_{max}$.
This explains the 
very weak dependence of $\alpha(\kappa)$ v.s. $\kappa$.

We conclude from the above discussion that,  
even if the PS coupling does not formally need any
regularization to insure its
stability, calculations without form factors -- though
analytically understood 
-- lead to results which are hardly interpretable on the
physical point of vue.

\vspace{0.2cm}
{\bf Acknowledgements:}
The numerical calculations were performed
at CGCV (CEA Grenoble) and  IDRIS (CNRS).
We are grateful to the staff members
of these two organizations for their constant support.


\begin{thebibliography}{9}
\bibitem{karm76}  V.A. Karmanov, ZhETF 71 (1976) 399 
(transl.: JETP  44 (1976) 210).
\bibitem{BPP_PR_98}   S.J. Brodsky, H.-C. Pauli and S.S. Pinsky,
Phys. Rep., 
{\bf 301} (1998) 299. 
\bibitem{These_MMB} M. Mangin-Brinet, {\it Th\`ese Universit\'e
de Paris} (2001).
\bibitem{fermions} M. Mangin-Brinet, J. Carbonell and V.A.
Karmanov, 
submitted for publication.
\bibitem{cdkm} J. Carbonell, B. Desplanques, V.A. Karmanov and 
J.-F. Mathiot, Phys. Reports 300 (1998) 215.
\bibitem{ckj0}    J. Carbonell and V.A. Karmanov, 
Nucl. Phys.  A589 (1995) 713.
\bibitem{ll} L.D. Landau, E.M. Lifshits, {\it Quantum
mechanics}, \S 35, Pergamon press, 1965. 
\bibitem{mck_prd1} M. Mangin-Brinet, J. Carbonell and V.A.
Karmanov, 
                   Phys Rev D  64 (2001) 027701.
\bibitem{mck_prd2} M. Mangin-Brinet, J. Carbonell and V.A.
Karmanov, 
                   accepted in Phys Rev D (2001).
\end{thebibliography}
\end{document}